\documentclass [10pt]{article}
\usepackage{graphicx}
\usepackage{dcolumn}
\usepackage{bm}
\usepackage {amssymb}
\usepackage {amsmath}
\usepackage {longtable}

\begin{document}


\title{Vortex\ structures\ with\ complex\ points singularities in the
two-dimensional \ Euler \ equation. New\ exact \ solutions.}

\maketitle
\author{Anatoly TUR, Vladimir YANOVSKY$^{\ast}$, Konstantin KULIK$^{\ast}$ }

\textit{ Universit\'{e} de Toulouse [UPS],\\
CNRS,Center d'Etude Spatiale des Rayonnements,\\9 avenue du Colonel Roche, BP 4346, \\
31028 Toulouse Cedex 4, France.
 E-mail address: anatoly.tour@cesr.fr}

 \textit{$^{\ast}$ Institute for Single Crystals, Nat. Academy of
Science Ukraine,Kharkov 31001, Lenin Ave.60, s, Ukraine\\
E-mail address: yanovsky@isc.kharkov.ua, koskul@isc.kharkov.ua}

\textit{keyword: 2D-Euler equations, exact solutions}


\textit{Physica D: Nonlinear Phenomena, Volume 240, Issue 13, p.
1069-1079.}

\begin{abstract}
In this work we found the new class of exact stationary solutions for
2D-Euler equations. Unlike of already known solutions, the new one contain
complex singularities. We consider as complex, point singularities which
have the vector field index greater than one. For example, the dipole
singularity is complex because its index is equal to two. We present in
explicit form a large class of exact localized stationary solutions for
2D-Euler equations with the singularity which index is equal to three. The
obtained solutions are expressed in terms of elementary functions. These
solutions represent complex singularity point surrounded by vortex
satellites structure. We discuss also motion equation of singularities and
conditions for singularity point stationarity which provides the
stationarity of complex vortex configuration.
\end{abstract}

%
%
%
%
%
%
%

\section{\protect\bigskip Introduction}

The importance of exact solutions for 2D-Euler equations is well
known. Today, the list of exact solutions is quite impressive.
Without pretending to be exhaustive we will mention only some of
them. First of all, this are classical solutions with smooth
vorticity, such as Rankine and Kirchhoff type vortices (see, for
example, the standard references \cite{[1]}- \cite {[3]} ),
elliptical Moore and Suffman vortices.\cite{[2]}, Lamb dipole
\cite {[1]} and Stuart vortex pattern \cite{[4a]}. It is
interesting, that these classical solutions are still topical (see
for example recent works \cite {[5]},\cite{[6]}). The
generalization of these solutions are the models of different
coherent structures, vortex patches, and vortex crystals (see, for
example, \cite{[7]}-\cite{[13]} and references therein), which are
well observed in numerical and laboratory experiments (see, for
example,\cite {[14]}-\cite{[21]}). Let us note that Stuart
solution is based on the equation of Liouville type for stream
function. Others classes of exact solutions rest upon the equation
of Sinh-Poisson type for the stream function \cite{[22]}-
\cite{[24]}. Some exact solutions in Lagrange coordinates are
given on the work \cite{[25]}. It should be noted the interesting
class of Kida -Neu vortices \cite{[26]},\cite{[27]}. Also there is
a lot of solutions with singular distribution of vorticity (see.
\cite {[2]} ), and solutions, which contain a smooth part of
vorticity field as well as point singularities \cite{[28]}-
\cite{[32]}, which are usually forming symmetrical configurations.
\ Numerous publications deal with particular cases of point
vortices (main references can be found in \cite
{[2]},\cite{[3]},\cite{[33]} and in works \cite{[34]}- \cite{[38]}
). It is known, that point vortices generate integrable dynamical
systems as well as dynamical systems with chaotical behavior (see,
for example,\cite{[3]}, \cite {[34]}, \cite{[37]},\cite{[38]} ).
The case of a couple of point vortices, which vorticities are of\
same magnitude but opposite signs arouses particular interest.
Such point vortices couples are usually called point dipoles. Many
works are dedicated to the dynamics and statistic of point dipoles
of this type (see, for example,\cite{[3]},\cite{[36]},\cite{[39]}-
\cite{[41]}.) Such point dipoles attract interest, especially in
the theory of equations Charney-Hasegava-Mima for atmosphere and
plasma. It means solutions of modon type \cite{[42]} and its
models obtained with the help of dipole structures (see, for
example,\cite{[43]}- \cite{[46a]}). Such a point dipole has two
elliptical singular points (Fig.\ref{Fig1}). However, there are
also 2D-point dipoles, which have different structure. They have
one complex singular point which has the vector field index
$\left| J\right| =2.$ This type of point dipoles appears naturally
as a consequence of expansion of stream function $\Psi $ in
multipoles moments, similar to usual electrodynamics. More
precisely, we suppose, that the vorticity $\omega $ is
localized in the restricted neighborhood $G$ \ of the \ point $%
\overrightarrow{r}_{0}$, which belongs to 2D-plane. Let us denote
the \ characteristic scale of \ domain $G$ by $a$, i.e.$\left\|
G\right\| \lesssim a.$ Then the stream function $\Psi$:

\begin{figure}
  \centering
  \includegraphics[width=6 cm]{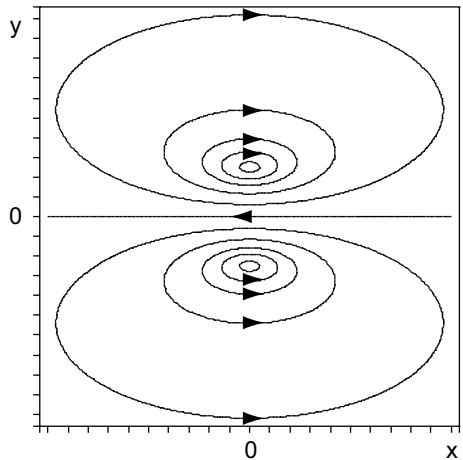}\\
  \caption{Couple of point vortices with vorticity of opposite signs,
  which is considered as point dipole.}\label{Fig1}
\end{figure}

\begin{equation}
\Psi (\overrightarrow{r},t)=-\frac{1}{4\pi }\int_{-\infty }^{+\infty }\omega
(\overrightarrow{a},t)\ln \left| \overrightarrow{r}-\overrightarrow{a}%
\right| ^{2}d\overrightarrow{a},  \label{1.1}
\end{equation}

in the domain $\left| \overrightarrow{r}\right| \gg \left| \overrightarrow{a}%
\right| ,$we can expand, as usual, in multipoles moments :

\begin{equation}
\Psi =-\frac{\Gamma _{0}}{2\pi }\ln \left| \overrightarrow{r}-%
\overrightarrow{r}_{0}\right| -\frac{1}{2\pi }\overrightarrow{D}\frac{%
\partial }{\partial \overrightarrow{x}}\ln \left| \overrightarrow{r}-%
\overrightarrow{r}_{0}\right| +\cdots  \label{1.2}
\end{equation}

Here, the first term is obviously point vortex with the vorticity
strength $\Gamma _{0}:\Gamma _{0}=\int \omega
(\overrightarrow{a})d\overrightarrow{a}.$ The second term is the
dipole $\Psi _{D}:$

\begin{equation*}
\Psi _{D}=\frac{1}{2\pi }\frac{\overrightarrow{D}(\overrightarrow{r}-%
\overrightarrow{r}_{0})}{\left| \overrightarrow{x}-\overrightarrow{r}%
_{0}\right| ^{2}},
\end{equation*}

 with dipole moment $\overrightarrow{D}:\overrightarrow{D}=\int
\omega (\overrightarrow{a})\overrightarrow{a}d\overrightarrow{a}.$
Further, we will call this dipole by dipole singularity to avoid
any confusion with point dipole which is composed of two point
vortices. It is easy to see the difference between them when
calculating the vorticity. Apparently, the
vorticity of vortices couple is proportional to the difference of two $%
\delta -$functions, whereas the vorticity of dipole singularity is
proportional to derivative of $\delta -$function:

\begin{equation*}
\omega \sim \delta ^{^{\prime }}(x)=\underset{h\rightarrow 0}{\lim }\frac{1}{%
h}[\delta (x+h)-\delta (x)],
\end{equation*}

However, if the parameter $h$ is considered as minor, but final,

\begin{equation*}
\omega \sim \frac{1}{h}[\delta (x+h)-\delta (x)],
\end{equation*}

then such a couple of closely located point vortices with the opposite
vorticity may be considered as a approximative model of dipole singularity.

The important role of dipole singularities in dynamic of 2D-Euler equation
was mentioned in the paper \cite{[47]}. It is shown in this paper, that
point dipole singularities are moving singularities of 2D- Euler equation
which are forming together with point vortices the hamiltonian system of
singulaties motion equation. This dynamical system has three independent
integrals of motion in involution of Kirchhoff type. This fact means the
complete integrability of the problem of motion of one dipole singularity
and of one point vortex.\ Corresponding \ exact solutions for the plane case
without boundaries are given in the paper \cite{[48]}. As it was noted
earlier, the dipole singularity is more complex than point vortex since the
index of its vector field is equal to two ( unlike simple singularities,
which index is equal to $\pm 1$). In this paper we will continue the studies
of solutions for\ 2D-Euler equation with complex singularities, which were
started in \cite{[47]}, \cite{[48]}. Also, we discuss in details some
questions which previously were presented briefly. Further, we demonstrate,
that more complex multipole singularities, generally speaking, are not
compatible with the dynamic of 2D- Euler equation. But in this work we show
that 2D-Euler equation has a new class of exact stationary solutions with
complex singular point, which index is equal to three. These solutions are
found in explicit form and expressed in terms of elementary functions.
Obtained solutions describe localized vortex structures, in which complex
singular point is surrounded by vortices satellites. In addition we discuss
the equation of motion for singularities and we give the sufficient
conditions of immobility for singular points; without this equation one can
not guarantee the stationarity of solution.

\bigskip

\section{Dipol singularities in 2D-Euler equation}

\bigskip We use 2D Euler equation in form of Poisson brackets for the
vorticity $\omega $ and the stream function $\Psi :$

\begin{equation}
\omega =-\Delta \Psi  \label{1}
\end{equation}

\begin{equation}
\frac{\partial \Delta \Psi }{\partial t}+\left\{ \Delta \Psi ,\Psi \right\}
=0,  \label{2}
\end{equation}

or in the form:

\begin{equation}
\frac{\partial \Delta \Psi }{\partial t}+V_{p}\frac{\partial \Delta \Psi }{%
\partial x_{p}}=\frac{d\Delta \Psi }{dt}=0.  \label{3}
\end{equation}

here $\overrightarrow{V}$- is fluid velocity. Poisson brackets has the usual
form:

\begin{equation*}
\left\{ A,B\right\} =\varepsilon _{ik}\frac{\partial A}{\partial x_{i}}\frac{%
\partial B}{\partial x_{k}}=\frac{\partial A}{\partial x}\frac{\partial B}{%
\partial y}-\frac{\partial A}{\partial y}\frac{\partial B}{\partial x}
\end{equation*}

\bigskip $\varepsilon _{ik}-$is single anti symmetrical tensor:

\begin{equation*}
\varepsilon _{12}=1,\varepsilon _{21}=-1,\varepsilon _{11}=\varepsilon
_{22}=0;
\end{equation*}

$\overrightarrow{x}=(x,y)=(x_{1},x_{2}).$ Velocity field $\overrightarrow{V}%
=(V_{x},V_{y})=(V_{1},V_{2})$ is written in the form :

\begin{equation}
V_{i}=\varepsilon _{ik}\frac{\partial \Psi }{\partial x_{k}}.  \label{4}
\end{equation}

As a stationary solutions for the Euler equation (\ref{2}) one use
often the \textit{anzatz:}

\begin{equation}
\Delta \Psi =f(\Psi ),  \label{5}
\end{equation}

where $f(\Psi )$ is arbitrary enough differential function of $\Psi .$ For
example, for the Lamb solution \cite{[1]}, $f(\Psi )$ is chosen as a linear
function, for the Stuart solution \cite{[4a]} $f(\Psi )=\exp (-\Psi )$ and
for solutions \cite{[22]} $f(\Psi )=-\sinh \Psi $.

In solutions of type (\ref{5}) the vorticity $\omega $ is smooth enough
function. There are also stationary solutions in which the vorticity has a
smooth part as well as one dimensional immobile singularities of point
vortex type ( see, for example, \cite{[28]}-\cite{[32]} ).

In order to find stationary solutions for Euler equation (\ref{2}) with
singularities of point vortex type it was proposed in the work \cite{[32]}
the \textit{anzatz:}

\begin{equation*}
\Delta \Psi =f(\Psi )-\sum_{\alpha =1}^{N}\Gamma ^{\alpha }\delta (%
\overrightarrow{x}-\overrightarrow{x}^{\alpha })
\end{equation*}

For the simplest case exact stationary solutions were found independently
and in a different way in papers \cite{[31]}, \cite{[32]}. In this work we
study a more general \textit{anzatz. }Let us suppose that the vorticity
\textit{\ }$\omega =-\Delta \Psi $ can be presented in the form:

\begin{equation}
\Delta \Psi =f(\Psi )-\sum_{\alpha =1}^{N}\Gamma ^{\alpha }\delta (%
\overrightarrow{x}-\overrightarrow{x}^{\alpha })-\sum_{\beta
=1}^{M}D_{i}^{\beta }\frac{\partial }{\partial x_{i}}\delta (\overrightarrow{%
x}-\overrightarrow{x}^{\beta }).  \label{9}
\end{equation}

In the expansion (\ref{9}) the coordinates of singularities $\overrightarrow{%
x}^{\alpha },\overrightarrow{x}^{\beta }$ and coefficients
$D_{i}^{\beta }$ may depend on time. Then it is supposed that
vorticity is composed of smooth part and singular part as well,
which is a generalized function. To begin, let us consider
physical sense of development (\ref{9}). First term in right part
is obviously a smooth part of vorticity field. Second group of
terms is vorticity, which is generated by the set of point
vortices $N$, with stream function $\Psi _{N}$:

\begin{equation*}
\Psi _{N}=-\frac{1}{4\pi }\sum_{\alpha =1}^{N}\Gamma ^{\alpha }\ln
\left\vert \overrightarrow{x}-\overrightarrow{x}^{\alpha }\right\vert ^{2}.
\end{equation*}

The third group of terms in development (\ref{9}) with first derivatives of $%
\delta $ - function describes the vorticity which is generated by
group of \ dipole singularities $M$:

\begin{equation*}
\Delta \Psi _{M}=-\sum_{\beta =1}^{M}D_{i}^{\beta }\frac{\partial }{\partial
x_{i}}\delta (\overrightarrow{x}-\overrightarrow{x}^{\beta }).
\end{equation*}

Actually, the let us apply the operator

\begin{equation}
D_{i}^{\beta }\frac{\partial }{\partial x_{i}}=D_{1}^{\beta }\frac{\partial
}{\partial x_{1}}+D_{2}^{\beta }\frac{\partial }{\partial x_{2}}  \label{10}
\end{equation}

to Laplace equation:

\begin{equation*}
\frac{1}{4\pi }\Delta \ln \left\vert \overrightarrow{x}-\overrightarrow{x}%
^{\beta }\right\vert ^{2}=\delta (\overrightarrow{x}-\overrightarrow{x}%
^{\beta })
\end{equation*}

It is obvious that:

\begin{equation}
\frac{1}{4\pi }\Delta \left(D_{i}^{\beta }\frac{\partial
}{\partial x_{i}}\right)\ln \left|
\overrightarrow{x}-\overrightarrow{x}^{\beta }\right|
^{2}=\left(D_{i}^{\beta }\frac{\partial }{\partial x_{i}}\right)\delta (\overrightarrow{%
x}-\overrightarrow{x}^{\beta }).  \label{11}
\end{equation}

That is why the vorticity which is generated by the third group of terms has
the stream function $\Psi _{M}:$

\begin{equation}
\Psi _{M}=-\frac{1}{4\pi }\sum_{\beta =1}^{M}\left(D_{i}^{\beta }\frac{\partial }{%
\partial x_{i}}\right)\ln \left| \overrightarrow{x}-\overrightarrow{x}^{\beta
}\right| ^{2},  \label{12}
\end{equation}

i.e.that is the sum $M$ stream functions of dipole singularities
in the form:

\begin{equation}
\Psi _{D}^{\beta }=\frac{1}{2\pi }D_{i}^{\beta }\frac{(x_{i}-x_{i}^{\beta })%
}{\left\vert \overrightarrow{x}-\overrightarrow{x}_{\beta }\right\vert ^{2}},
\label{15}
\end{equation}

which are in the points $\overrightarrow{x}_{\beta }$ and have dipole
moments $\frac{1}{2\pi }\overrightarrow{D}^{\beta }$. The velocity field of
dipole singularity has evidently the form:

\begin{equation*}
V_{x}^{D}=-\frac{\partial \Psi _{D}}{\partial y}=-\frac{1}{2\pi \left|
\overrightarrow{x}-\overrightarrow{x}_{0}\right| ^{2}}\left[ D_{y}-\frac{2y(%
\overrightarrow{D}\overrightarrow{x})}{\left| \overrightarrow{x}-%
\overrightarrow{x}_{0}\right| ^{2}}\right] ,
\end{equation*}

\begin{equation*}
V_{y}^{D}=\frac{\partial \Psi _{D}}{\partial x}=\frac{1}{2\pi \left|
\overrightarrow{x}-\overrightarrow{x}_{0}\right| ^{2}}\left[ D_{x}-\frac{2x(%
\overrightarrow{D}\overrightarrow{x})}{\left| \overrightarrow{x}-%
\overrightarrow{x}_{0}\right| ^{2}}\right] .
\end{equation*}

The presence of complementary sources of vorticity in form of derivatives of
$\delta $ - function, itself, is not forbidden in (\ref{9}), if they are
compatible with Euler equation (\ref{2}). We have to remind \ that the
singularity part of vorticity like every generalized function with point
support is composed of $\delta -$ function sum and its derivatives only.
However, in the next chapter we show that the condition of compatibility of
development (\ref{9}) with Euler equation is not trivial and engenders
important restrictions for (\ref{9}). Formulae which one needs to work with
derivatives of generalized functions are well known, but for reader's
convenience we gave them in the Appendix. \ Before studying the general
expansion of vorticity (\ref{9}) let us examine \ some particular cases. We
begin with trivial solution, which is describing \ one point vortex. Let us
remind, how from point of view of generalized functions theory, point vortex
satisfies formally Euler equation. We will substitute vorticity and stream
function of point vortex which are in point $\overrightarrow{x}_{0}=0,$ in
Poisson brackets. Then we obtain:

\begin{equation}
\left\{ \Psi ,\Delta \Psi \right\} =\frac{\Gamma ^{2}}{2\pi (x^{2}+y^{2})}[%
x\delta (x)]\delta ^{^{\prime }}(y)-\frac{\Gamma ^{2}}{2\pi (x^{2}+y^{2})}[%
y\delta (y)]\delta ^{^{\prime }}(x)  \label{16}
\end{equation}

This Poisson brackets is equal to zero since the terms in brackets are equal
to zero from the generalized function theory point of view. The fact that
Poisson brackets is equal to zero (\ref{16}) is physically interpreted as
absence of self-interaction in point vortex.If the Poisson bracket does not
vanish for the solution with one singularity on the plane, then, typically,
a self acceleration of this singularity appears . This effect is considered
unacceptable from a physical point of view and the corresponding solutions
must be rejected.

Let us examine in the same way one dipole singularity, which is in the point
$\overrightarrow{x}_{0}$ and has the dipole moment $\overrightarrow{D}.$

The Poisson brackets (\ref{2}) for one dipole singularity gets the
form:

\begin{equation}
\left\{ \Psi _{D},\Delta \Psi _{D}\right\} =\frac{\partial \Psi _{D}}{%
\partial x}\frac{\partial }{\partial y}\overrightarrow{D}\frac{\partial }{%
\partial \overrightarrow{x}}\delta (\overrightarrow{x}-\overrightarrow{x}%
_{0})-\frac{\partial \Psi _{D}}{\partial y}\frac{\partial }{\partial x}%
\overrightarrow{D}\frac{\partial }{\partial \overrightarrow{x}}\delta (%
\overrightarrow{x}-\overrightarrow{x}_{0}).  \label{17}
\end{equation}

Let us show that Poisson brackets (\ref{17}) is equal to zero. We
substitute in (\ref{17}) derivatives in explicit form and
calculate it in details:

\begin{equation}
\frac{\partial }{\partial x}\overrightarrow{D}\frac{\partial }{\partial
\overrightarrow{x}}\delta (\overrightarrow{x}-\overrightarrow{x}%
_{0})=D_{1}\delta _{xx}^{^{\prime \prime }}(x-x_{0})\delta
(y-y_{0})+D_{2}\delta _{x}^{^{\prime }}(x-x_{0})\delta _{y}^{^{\prime
}}(y-y_{0}),  \label{18}
\end{equation}

\begin{equation}
\frac{\partial }{\partial y}\overrightarrow{D}\frac{\partial }{\partial
\overrightarrow{x}}\delta (\overrightarrow{x}-\overrightarrow{x}%
_{0})=D_{1}\delta _{x}^{^{\prime }}(x-x_{0})\delta _{y}^{^{\prime
}}(y-y_{0})+D_{2}\delta (x-x_{0})\delta _{yy}^{^{\prime \prime }}(y-y_{0}).
\label{19}
\end{equation}

\begin{equation}
\frac{\partial \Psi _{D}}{\partial x}=\frac{1}{2\pi }\frac{%
D_{1}[(y-y_{0})^{2}-(x-x_{0})^{2}]-2D_{2}(x-x_{0})(y-y_{0})}{\left|
\overrightarrow{x}-\overrightarrow{x}_{0}\right| ^{4}},  \label{20}
\end{equation}

\begin{equation}
\frac{\partial \Psi _{D}}{\partial y}=\frac{1}{2\pi }\frac{%
D_{2}[(x-x_{0})^{2}-(y-y_{0})^{2}]-2D_{1}(x-x_{0})(y-y_{0})}{\left|
\overrightarrow{x}-\overrightarrow{x}_{0}\right| ^{4}}.  \label{21}
\end{equation}

As a result we obtain:

\begin{equation}
2\pi \left\{ \Psi _{D},\Delta \Psi _{D}\right\} =\left[\frac{%
(D_{1}^{2}+D_{2}^{2})(y-y_{0})^{2}}{\left| \overrightarrow{x}-%
\overrightarrow{x}_{0}\right| ^{4}}\delta _{y}^{^{\prime
}}(y-y_{0})\right]\delta _{x}^{^{\prime }}(x-x_{0})-  \label{22}
\end{equation}

\begin{equation*}
-\left[\frac{(D_{1}^{2}+D_{2}^{2})(x-x_{0})^{2}}{\left| \overrightarrow{x}-%
\overrightarrow{x}_{0}\right| ^{4}}\delta _{x}^{^{\prime
}}(x-x_{0})\right]\delta _{y}^{^{\prime }}(y-y_{0})+
\end{equation*}

\begin{equation*}
+\frac{D_{1}D_{2}[(y-y_{0})^{2}-(x-x_{0})^{2}]-2D_{2}^{2}(x-x_{0})(y-y_{0})}{%
\left| \overrightarrow{x}-\overrightarrow{x}_{0}\right| ^{4}}\delta
(x-x_{0})\delta _{yy}^{^{\prime \prime }}(y-y_{0})-
\end{equation*}

\begin{equation*}
-\frac{D_{1}D_{2}[(x-x_{0})^{2}-(y-y_{0})^{2}]-2D_{1}^{2}(x-x_{0})(y-y_{0})}{%
\left| \overrightarrow{x}-\overrightarrow{x}_{0}\right| ^{4}}\delta
(y-y_{0})\delta _{xx}^{^{\prime \prime }}(x-x_{0}).
\end{equation*}

Let us use formulae from the Appendix. Action of first derivative of $\delta
-$function on usual functions is given by formula (\ref{A.2}). It follows
from this formula that the first and second terms in formula (\ref{22}) are
zero, since they contain zeros of the following form:

\begin{equation*}
(y-y_{0})\delta (y-y_{0});(y-y_{0})^{2}\mid _{y-y_{0}=0}\delta
_{y}^{^{\prime }}(y-y_{0});
\end{equation*}

\begin{equation*}
(x-x_{0})\delta (x-x_{0});(x-x_{0})^{2}\mid _{x-x_{0}=0}\delta
_{x}^{^{\prime }}(x-x_{0}).
\end{equation*}

Besides, in third and fourth terms the term with the factor
$D_{2}^{2},$ turns into zero because it contains zeros
$(x-x_{0})\delta (x-x_{0})$, and the term with the factor
$D_{1}^{2}$ turns into zero in the same way, because it contains
zero $(y-y_{0})\delta (y-y_{0}).$ In addition, it is evident that
the terms: $D_{1}D_{2}[(x-x_{0})^{2}\delta (x-x_{0})]\delta
_{yy}^{^{\prime \prime }}(y-y_{0})$ and

$D_{1}D_{2}[(y-y_{0})^{2}\delta (y-y_{0})]\delta _{xx}^{^{\prime \prime
}}(x-x_{0})$ are equal to zero. Consequently the brackets (\ref{22}) is
equal to:

\begin{equation}
2\pi \left\{ \Psi _{D},\Delta \Psi _{D}\right\} =\frac{%
D_{1}D_{2}(y-y_{0})^{2}}{\left| \overrightarrow{x}-\overrightarrow{x}%
_{0}\right| ^{4}}\delta (x-x_{0})\delta _{yy}^{^{\prime \prime }}(y-y_{0})-%
\label{23}
\end{equation}
\[-\frac{D_{1}D_{2}(x-x_{0})^{2}}{\left| \overrightarrow{x}-\overrightarrow{x}%
_{0}\right| ^{4}}\delta (y-y_{0})\delta _{xx}^{^{\prime \prime
}}(x-x_{0})\]

Now, for computing this equation we use the formula (\ref{A.4}),
which is needed to apply the second derivative of $\delta
-$function in the commutator (\ref{23}). It is obvious that all
the terms turn into zero, excluding terms without derivatives of
$\delta -$function, which are mutually eliminated:

\begin{equation*}
2\pi \left\{ \Psi _{D},\Delta \Psi _{D}\right\} =\frac{2D_{1}D_{2}}{\left|
\overrightarrow{x}-\overrightarrow{x}_{0}\right| ^{4}}\delta (%
\overrightarrow{x}-\overrightarrow{x}_{0})-\frac{2D_{1}D_{2}}{\left|
\overrightarrow{x}-\overrightarrow{x}_{0}\right| ^{4}}\delta (%
\overrightarrow{x}-\overrightarrow{x}_{0})=0.
\end{equation*}

\begin{figure}
  \centering
  \includegraphics[width=6 cm]{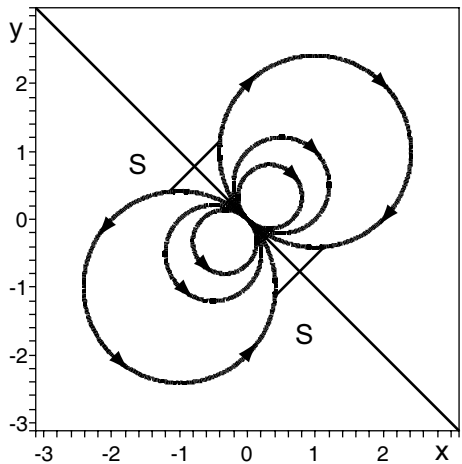}\\
  \caption{Dipole singularity with the index of vector field equal to 2.}\label{Fig2}
\end{figure}

Thereby we proved that there is no self-interaction in dipole
singularity and, consequently, it is the exact stationary solution
of Euler equation. We can also understand the absence of
self-interaction in dipole singularity basing on simple physical
considerations. Actually, from (Fig.\ref{Fig2}) one can see, that
due to symmetry of stream line configuration, the flux of impulse
which is flowing in singularity through arbitrary section $S$ is
exactly equal to impulse flux which is flowing out from
singularity through the same symmetrical section $S$. That means
that the force which is acting on singularity is equal to zero and
the singularity does not move.

Naturally the question arises: do the higher multipole
singularities, for example, quadruple, self-interaction? To answer
this question we have to calculate Poisson brackets $\left\{ \Psi
,\Delta \Psi \right\} $ for \ quadruple singularity. The quadruple
singularity has the stream function:

\begin{equation}
\Psi _{D_{2}}=-\frac{1}{8\pi }D_{i_{1}i_{2}}\frac{\partial ^{2}}{\partial
x_{i_{1}}\partial x_{i_{2}}}\ln \left\vert \overrightarrow{x}-%
\overrightarrow{x}_{0}\right\vert ^{2},  \label{23.1}
\end{equation}

and the vorticity:

\begin{equation}
\omega =\frac{1}{2}D_{i_{1}i_{2}}\frac{\partial ^{2}}{\partial
x_{i_{1}}\partial x_{i_{2}}}\delta (\overrightarrow{x}-\overrightarrow{x}%
_{0}).  \label{23.2}
\end{equation}

When computing the Poisson brackets $\left\{ \Psi ,\Delta \Psi \right\} $
using the formula (\ref{A.4.1}) \ we obtain the following result:

\begin{equation}
\left\{ \Psi _{D_{2}},\Delta \Psi _{D_{2}}\right\} =-\frac{1}{3\pi }\frac{%
D_{12}(D_{11}+D_{22})}{(x^{2}+y^{2})^{3}}\delta (\overrightarrow{x}),
\label{23.3}
\end{equation}

which shows that there is self-interaction in the general case in quadruple
singularity; (exception to the rule is special choice of coefficients: $%
D_{12}=0,$ or $D_{11}+D_{22}=0$). The self-interaction means that
the given singularity is not physical. The similar result is
obtained for multipoles of higher orders. Therefore, in general
case the multipoles singularities of orders higher than dipole are
not compatible with Euler equation.

\section{Singularities motion and stationarity conditions.}

According to the work \cite{[47]} let us examine the question in which cases
vorticity expansion (\ref{9}) is compatible with Euler equation (\ref{2})
and which restrictions arise from it \ for stationary and not stationary
cases. We consider singularities coordinates $\overrightarrow{x}_{\alpha },%
\overrightarrow{x}_{\beta }$ and all coefficients $\Gamma ^{\alpha
},D_{i}^{\beta }$ for general case as depending on time. Further, we will
distinguish two cases: $f(\Psi )=0,f(\Psi )\neq 0.$ Let us first examine the
first case $f(\Psi )=0.$ In this case the smooth part of vorticity is absent
and the stream function $\Psi $ is determined by singularities only

\begin{equation}
\Psi =\Psi _{N}+\Psi _{M}  \label{24}
\end{equation}

and derivatives of vorticity $\Delta \Psi $ obviously have the form :

\begin{equation}
\frac{\partial \Delta \Psi }{\partial t}=-\sum_{\alpha =1}^{N}\frac{d\Gamma
^{\alpha }}{dt}\delta (\overrightarrow{x}-\overrightarrow{x}^{\alpha
})+\sum_{\alpha =1}^{N}\Gamma ^{\alpha }\frac{\partial \delta (%
\overrightarrow{x}-\overrightarrow{x}^{\alpha })}{\partial x_{i}}\frac{%
dx_{i}^{\alpha }}{dt}-  \label{25}
\end{equation}

\begin{equation*}
-\sum_{\beta =1}^{M}\frac{dD_{i}^{\beta }}{dt}\frac{\partial \delta (%
\overrightarrow{x}-\overrightarrow{x}^{\beta })}{\partial x_{i}}+\sum_{\beta
=1}^{M}D_{i}^{\beta }\frac{\partial ^{2}\delta (\overrightarrow{x}-%
\overrightarrow{x}^{\beta })}{\partial x_{i}\partial x_{k}}\frac{%
dx_{k}^{\beta }}{dt},
\end{equation*}

\begin{equation}
V_{p}\frac{\partial \Delta \Psi }{\partial x_{p}}=-V_{p}\sum_{\alpha
=1}^{N}\Gamma ^{\alpha }\frac{\partial \delta (\overrightarrow{x}-%
\overrightarrow{x}^{\alpha })}{\partial x_{p}}-V_{p}\sum_{\beta
=1}^{M}D_{i}^{\beta }\frac{\partial ^{2}\delta (\overrightarrow{x}-%
\overrightarrow{x}^{\beta })}{\partial x_{i}\partial x_{p}}.  \label{26}
\end{equation}

Where $V_{p}$ - is fluid velocity (\ref{4}), which can be written
down in the following form:

\begin{equation}
V_{p}=-\frac{1}{4\pi }\varepsilon _{pq}\frac{\partial }{\partial x_{q}}\left[%
\sum_{\alpha =1}^{N}\Gamma ^{\alpha }\ln \left| \overrightarrow{x}-%
\overrightarrow{x}^{\alpha }\right| ^{2}+\sum_{\beta =1}^{M}(D_{i}^{\beta }%
\frac{\partial }{\partial x_{i}})\ln \left| \overrightarrow{x}-%
\overrightarrow{x}^{\beta }\right| ^{2}\right].  \label{27}
\end{equation}
We substitute (\ref{25}) and (\ref{26}) in Euler equation(\ref{3}), and we
obtain:

\begin{equation}
\sum_{\alpha =1}^{N}\left[\frac{d\Gamma ^{\alpha }}{dt}\delta (\overrightarrow{x}-%
\overrightarrow{x}^{\alpha })-\Gamma ^{\alpha }\frac{\partial \delta (%
\overrightarrow{x}-\overrightarrow{x}^{\alpha })}{\partial x_{i}}\frac{%
dx_{i}^{\alpha }}{dt}+V_{p}\Gamma ^{\alpha }\frac{\partial \delta (%
\overrightarrow{x}-\overrightarrow{x}^{\alpha })}{\partial
x_{p}}\right]+ \label{28}
\end{equation}

\begin{equation*}
+\sum_{\beta =1}^{M}\left[\frac{dD_{i}^{\beta }}{dt}\frac{\partial \delta (%
\overrightarrow{x}-\overrightarrow{x}^{\beta })}{\partial x_{i}}%
-D_{i}^{\beta }\frac{\partial ^{2}\delta (\overrightarrow{x}-\overrightarrow{%
x}^{\beta })}{\partial x_{i}\partial x_{k}}\frac{dx_{k}^{\beta }}{dt}%
+V_{p}D_{i}^{\beta }\frac{\partial ^{2}\delta (\overrightarrow{x}-%
\overrightarrow{x}^{\beta })}{\partial x_{i}\partial
x_{p}}\right]=0.
\end{equation*}

Since equations for different singularities $\overrightarrow{x}^{\alpha }(t),%
\overrightarrow{x}^{\beta }(t)$ must go to zero independently, we obtain:

\begin{equation}
\sum_{\alpha =1}^{N}\left[\frac{d\Gamma ^{\alpha }}{dt}\delta (\overrightarrow{x}-%
\overrightarrow{x}^{\alpha })-\Gamma ^{\alpha }\frac{\partial \delta (%
\overrightarrow{x}-\overrightarrow{x}^{\alpha })}{\partial x_{i}}\frac{%
dx_{i}^{\alpha }}{dt}+V_{p}\Gamma ^{\alpha }\frac{\partial \delta (%
\overrightarrow{x}-\overrightarrow{x}^{\alpha })}{\partial
x_{p}}\right]=0, \label{29}
\end{equation}

\begin{equation}
\sum_{\beta =1}^{M}\left[\frac{dD_{i}^{\beta }}{dt}\frac{\partial \delta (%
\overrightarrow{x}-\overrightarrow{x}^{\beta })}{\partial x_{i}}%
-D_{i}^{\beta }\frac{\partial ^{2}\delta (\overrightarrow{x}-\overrightarrow{%
x}^{\beta })}{\partial x_{i}\partial x_{k}}\frac{dx_{k}^{\beta }}{dt}%
+V_{p}D_{i}^{\beta }\frac{\partial ^{2}\delta (\overrightarrow{x}-%
\overrightarrow{x}^{\beta })}{\partial x_{i}\partial
x_{p}}\right]=0. \label{30}
\end{equation}

Every equation (\ref{29}),(\ref{30}) contains also singularities
of different orders ($\delta -$functions and derivatives of
$\delta -$ function). To satisfy each of equation (\ref{29}),
(\ref{30}) the factors of singularities of different orders shall
also go to zero independently. Let
us begin with the simplest equation (\ref{29}). The derivative of $\delta -$%
function acts on velocity field $\overrightarrow{V}$ according to formula (%
\ref{A.2}). Because of non compressibility condition $\frac{\partial V_{p}}{%
\partial x_{p}}=0,$ the third term in formula (\ref{29}) does not give
additional contributions to singularities with $\delta -$functions. That is
why when factors of $\delta -$functions are going to zero this gives the
well known equation:

\begin{equation*}
\frac{d\Gamma ^{\alpha }}{dt}=0;\forall \alpha ,
\end{equation*}

i.e. conservation of vorticity strength. Terms with the first derivatives
from $\delta -$function give:

\begin{equation}
\sum_{\alpha =1}^{N}\Gamma ^{\alpha }\left(\frac{dx_{p}^{\alpha }}{dt}-V_{p}\right)%
\frac{\partial \delta (\overrightarrow{x}-\overrightarrow{x}^{\alpha })}{%
\partial x_{p}}=0,  \label{31}
\end{equation}

i.e. point vortices motion equations:

\begin{equation}
\left(\frac{dx_{p}^{\alpha }}{dt}-V_{p}\mid _{\overrightarrow{x}=\overrightarrow{x%
}^{\alpha }}\right)=0,(\alpha =1,2,...N)  \label{32}
\end{equation}

(In equations (\ref{32}) there are no terms with self-interaction). Let us
examine now equations for point dipoles (\ref{30}). The third term in
equation (\ref{30}) contains second derivatives from $\delta -$function.
These derivatives act on velocity field $V_{p}(\overrightarrow{x},t)$
according to formula (\ref{A.4}) in Appendix, and give the first ones and
second ones as well. Since derivatives of different orders must be set equal
to zero independently, the equation (\ref{30}) splits into two groups of
equations:

\begin{equation}
\sum_{\beta =1}^{M}\left(\frac{dD_{i}^{\beta }}{dt}-D_{i}^{\beta
}\frac{\partial
V_{p}}{\partial x_{i}}\mid _{\overrightarrow{x}=\overrightarrow{x}^{\beta }}\right)%
\frac{\partial \delta (\overrightarrow{x}-\overrightarrow{x}^{\beta })}{%
\partial x_{p}}=0,  \label{33}
\end{equation}

\begin{equation}
\sum_{\beta =1}^{M}D_{i}^{\beta }\left(\frac{dx_{p}^{\beta }}{dt}-V_{p}\mid _{%
\overrightarrow{x}=\overrightarrow{x}^{\beta }}\right)\frac{\partial ^{2}\delta (%
\overrightarrow{x}-\overrightarrow{x}^{\beta })}{\partial x_{i}\partial x_{p}%
}=0.  \label{34}
\end{equation}

Equations (\ref{33}),(\ref{34}) are satisfied, if all factors of different
singularities turn into zero independently and the number of equations
coincides with the number of variables. Terms of point dipole with
self-interaction as this was shown in previous chapter are absent. As a
result we obtain motion equation of point dipoles:

\begin{equation}
\frac{dD_{i}^{\beta }}{dt}=D_{i}^{\beta }\frac{\partial V_{p}}{\partial x_{i}%
}\mid _{\overrightarrow{x}=\overrightarrow{x}^{\beta }},  \label{35}
\end{equation}

\begin{equation}
\frac{dx_{p}^{\beta }}{dt}=V_{p}\mid _{\overrightarrow{x}=\overrightarrow{x}%
^{\beta }},(\beta =1,2,...M).  \label{36}
\end{equation}

From (\ref{33}),(\ref{34}) follows that if all dipole moments are
equal to zero, $(D_{i}^{\beta }=0,\forall \beta )$, then equations
(\ref{33}),(\ref {34}) are absent and only the system of equation
(\ref{32}) remains for point vortices motion. And vice versa, one
can see from equation (\ref{31})
that if all $\Gamma ^{\alpha }=0,\forall \alpha ,$ then the system (\ref{31}%
) is absent and only motion equations of point dipoles remain. For
the general case the equation system
(\ref{32}),(\ref{35}),(\ref{36}), where the velocity has the form
(\ref{27}), describes motion of point vortices $N$ as well as
motion of point dipoles $M$. This equation system was obtained for
the first time in \cite{[47]}, where, in particular, was shown,
that it can be written down in the Hamiltonian form. Further, to
write down the equation
system of singularities motion, we denote coordinates of point vortices as $%
x_{vi}^{\alpha }$ (where $\alpha -$ is the number of point vortex
$\alpha =(1,2,...N)$ , index $i$ takes the values $i=1,2,$
i.e.denotes the coordinates of point vortex $x_{1}\equiv
x,x_{2}\equiv y$, $\ $the subscript $v$ \ means the vortex
coordinates). In similar manner we denote the
coordinates of dipole singularities as $x_{di}^{\beta }$ $($where $\beta -$%
is the number of dipole singularity $\beta =1,2,...M).$ Then, motion
equation of singularities (\ref{32}),(\ref{35}),(\ref{36}), taking into
account formula (\ref{27}) for fluid velocity, takes the form of :
\begin{equation}
\frac{dx_{vi}^{\alpha }}{dt}=-\varepsilon
_{ik}\left[\sum\limits_{\gamma \neq \alpha }^{N}\frac{\Gamma
_{\gamma }}{2\pi }\frac{(x_{vk}^{\alpha
}-x_{vk}^{\gamma })}{\left| \overrightarrow{x}_{v}^{\alpha }-\overrightarrow{%
x}_{v}^{\gamma }\right| ^{2}}+ \right. \label{36.1}
\end{equation}

\begin{equation*}
\left.+\sum\limits_{\beta =1}^{M}\frac{D_{l}^{\beta }}{\pi }\left(\frac{\delta _{lk}}{%
\left\vert \overrightarrow{x}_{v}^{\alpha }-\overrightarrow{x}_{d}^{\beta
}\right\vert ^{2}}-\frac{2(x_{vl}^{\alpha }-x_{dl}^{\beta })(x_{vk}^{\alpha
}-x_{dk}^{\beta })}{\left\vert \overrightarrow{x}_{v}^{\alpha }-%
\overrightarrow{x}_{d}^{\beta }\right\vert ^{4}}\right)\right]=0,
\end{equation*}

\begin{equation}
\frac{dx_{di}^{\beta }}{dt}=-\varepsilon _{ik}\left[\sum\limits_{\alpha =1}^{N}%
\frac{\Gamma _{\alpha }}{2\pi }\frac{(x_{dk}^{\beta }-x_{vk}^{\alpha })}{%
\left\vert \overrightarrow{x}_{v}^{\alpha
}-\overrightarrow{x}_{d}^{\beta }\right\vert ^{2}}+ \right.
\label{36.2}
\end{equation}

\begin{equation*}
\left.+\sum\limits_{\gamma \neq \beta }^{M}\frac{D_{l}^{\gamma }}{\pi }\left(\frac{%
\delta _{lk}}{\left\vert \overrightarrow{x}_{d}^{\beta }-\overrightarrow{x}%
_{d}^{\gamma }\right\vert ^{2}}-\frac{2(x_{dl}^{\beta }-x_{dl}^{\gamma
})(x_{dk}^{\beta }-x_{dk}^{\gamma })}{\left\vert \overrightarrow{x}%
_{d}^{\beta }-\overrightarrow{x}_{d}^{\gamma }\right\vert
^{4}}\right)\right],
\end{equation*}

\begin{equation}
\frac{dD_{i}^{\beta }}{dt}=D_{m}^{\beta }\varepsilon
_{ik}\left[\sum\limits_{\alpha =1}^{N}\frac{\Gamma _{\alpha }}{2\pi }\left(\frac{%
\delta _{km}}{\left\vert \overrightarrow{x}_{d}^{\beta }-\overrightarrow{x}%
_{v}^{\alpha }\right\vert ^{2}}-\frac{2(x_{dk}^{\beta }-x_{vk}^{\alpha
})(x_{dm}^{\beta }-x_{vm}^{\alpha })}{\left\vert \overrightarrow{x}%
_{d}^{\beta }-\overrightarrow{x}_{d}^{\gamma }\right\vert
^{4}}\right)+\right. \label{36.3}
\end{equation}

\begin{equation*}
+\sum\limits_{\gamma \neq \beta }^{M}\frac{D_{l}^{\gamma }}{\pi }\left(-\frac{%
2\delta _{lk}(x_{dm}^{\beta }-x_{dm}^{\gamma })}{\left\vert \overrightarrow{x%
}_{d}^{\beta }-\overrightarrow{x}_{d}^{\gamma }\right\vert ^{4}}-\frac{%
2\delta _{ml}(x_{dk}^{\beta }-x_{dk}^{\gamma })}{\left\vert \overrightarrow{x%
}_{d}^{\beta }-\overrightarrow{x}_{d}^{\gamma }\right\vert
^{4}}-\right.
\end{equation*}

\begin{equation*}
\left.\left.-\frac{2\delta _{mk}(x_{dl}^{\beta }-x_{dl}^{\gamma
})}{\left\vert \overrightarrow{x}_{d}^{\beta
}-\overrightarrow{x}_{d}^{\gamma }\right\vert
^{4}}+\frac{8(x_{dl}^{\beta }-x_{dl}^{\gamma })(x_{dk}^{\beta
}-x_{dk}^{\gamma })(x_{dm}^{\beta }-x_{dm}^{\gamma })}{\left\vert
\overrightarrow{x}_{d}^{\beta }-\overrightarrow{x}_{d}^{\gamma
}\right\vert ^{6}}\right)\right].
\end{equation*}

Equations (\ref{36.1}), (\ref{36.2}) describe the motion of point vortices
and dipole singularities as well and equation (\ref{36.3}) describes time
evolution of dipole moment. As this is shown in work \cite{[47]} these
equations have the Hamiltonian form:

\bigskip

\begin{equation}
\Gamma _{\alpha }\frac{dx_{vi}^{\alpha }}{dt}=\varepsilon _{ik}\frac{%
\partial H}{\partial x_{vk}^{\alpha }},  \label{36.4}
\end{equation}

\begin{equation*}
\frac{dx_{di}^{\beta }}{dt}=-\varepsilon _{ik}\frac{\partial H}{\partial
D_{k}^{\beta }},
\end{equation*}

\begin{equation*}
\frac{dD_{i}^{\beta }}{dt}=-\varepsilon _{ik}\frac{\partial H}{\partial
x_{dk}^{\alpha }},
\end{equation*}

Here Hamiltonian $H$ has the form:

\begin{equation}
H=-\frac{1}{4\pi }\sum_{\substack{ \alpha =1,\beta =1  \\ \alpha \neq \beta
}}^{N}\Gamma _{\alpha }\Gamma _{\beta }\ln \left\vert \overrightarrow{x}%
_{v}^{\alpha }-\overrightarrow{x}_{v}^{\beta }\right\vert -\frac{1}{2\pi }%
\sum_{\alpha =1,\beta =1}^{\alpha =N,\beta =M}\frac{\Gamma _{\alpha
}D_{l}^{\beta }(x_{vl}^{\alpha }-x_{dl}^{\beta })}{\left\vert
\overrightarrow{x}_{v}^{\alpha }-\overrightarrow{x}_{d}^{\beta }\right\vert
^{2}}-  \label{36.5}
\end{equation}

\begin{equation*}
-\sum_{\substack{ \beta \prec \gamma =1  \\ \beta \neq \gamma }}^{M}\frac{%
2D_{m}^{\beta }(x_{dm}^{\beta }-x_{dm}^{\gamma })D_{l}^{\gamma
}(x_{dl}^{\beta }-x_{dl}^{\gamma })-D_{m}^{\beta }D_{m}^{\gamma }(%
\overrightarrow{x}_{d}^{\beta }-\overrightarrow{x}_{d}^{\gamma })^{2}}{\pi
\left\vert \overrightarrow{x}_{d}^{\beta }-\overrightarrow{x}_{d}^{\gamma
}\right\vert ^{4}}.
\end{equation*}

Equation system (\ref{36.4}) has the conservation laws of the type of
Kirchhoff's generalized integrals i.e. the conservation laws related to
motion equation invariance under translation and rotation of coordinates
system:

\begin{equation}
I_{1}=\sum_{\alpha =1}^{N}\Gamma _{\alpha }x_{v1}^{\alpha }-\sum_{\beta
=1}^{M}D_{1}^{\beta }=const.  \label{36.6}
\end{equation}

\begin{equation}
I_{2}=\sum_{\alpha =1}^{N}\Gamma _{\alpha }x_{v2}^{\alpha }-\sum_{\beta
=1}^{M}D_{2}^{\beta }=const.  \label{36.7}
\end{equation}

\begin{equation}
J=\sum_{\alpha =1}^{N}\Gamma _{\alpha }(\overrightarrow{x}_{v}^{\alpha
})^{2}-2\sum_{\beta =1}^{M}(\overrightarrow{D}^{\beta }\overrightarrow{x}%
_{d}^{\beta })=const.  \label{36.8}
\end{equation}

As in case of point vortices there are three independent motion
integrals in the involution: $H,J$ and $I_{1}^{2}+I_{2}^{2}$. This
means that the motion of one vortex and one point dipole are
integrable. The exact non stationary solutions for plane case
without boundaries are given in \cite{[48]}. Naturally, the
question arises, is it possible or not to add for the non
stationary case in expansion (\ref{9}) more higher derivatives of
$\delta $ - function, i.e.multipoles of higher order. Multipoles
of higher order, than dipole produce two kinds of difficulties.
First of all, as it was shown \ in sectoin 2, that, generally
speaking, singularities of this kind have self-interaction. From
dynamical point of view, the substitution of high multipoles in
Euler equation in accordance with the formula (\ref{A.3}) Appendix
A, engenders overdetermined equation system. Hence, for the non
stationary case, vorticity expansion (\ref{9}) is compatible with
Euler equation with $f(\Psi )=0,$ if the following sufficient
conditions are satisfied:

1.\qquad Multipole moments starting from the quadruple one are equal to zero.

2.\qquad Singularities motion obeys the equations (\ref{36.4}).

Let us now examine stationary case, when $f(\Psi )\neq 0,\frac{df(\Psi )}{%
d\Psi }\neq 0.$ In this case the stream function can depend on time since
singularities coordinates and dipoles moments are function on time. The
substitution of vorticity (\ref{9}) in Euler equation (\ref{3}) gives for
the smooth part of stream function the equation:

\begin{equation}
\left\lbrack \frac{\partial \Psi }{\partial t}+\sum_{\alpha =1}^{N}\left(\frac{%
\partial \Psi }{\partial x_{v}^{\alpha }}\frac{dx_{v}^{\alpha }}{dt}+\frac{%
\partial \Psi }{\partial y_{v}^{\alpha }}\frac{dy_{v}^{\alpha }}{dt}%
\right)+\sum_{\beta =1}^{M}\left(\frac{\partial \Psi }{\partial x_{d}^{\beta }}\frac{%
dx_{d}^{\beta }}{dt}+\frac{\partial \Psi }{\partial y_{d}^{\beta }}\frac{%
dy_{d}^{\beta }}{dt}\right)+ \right. \label{37}
\end{equation}

\begin{equation*}
\left.+\sum_{\beta =1}^{M}\frac{\partial \Psi }{\partial D_{i}^{\beta }}\frac{%
dD_{i}^{\beta }}{dt}\right]\frac{df(\Psi )}{d\Psi }=0.
\end{equation*}

The first term is related to explicit dependence of stream function $\Psi $
on time, the second one and the third one are related to singularities
motion. The fourth term in (\ref{37}) is related to dependence of dipole
moments on time.\ We have to add to the equation (\ref{37}) equations for
singularities parts of vorticity field which were already given (formulae (%
\ref{36.4})). In the case $\frac{df(\Psi )}{d\Psi }=0,$ the equation (\ref
{37}) is absent and only singularities motion equations remain. The
sufficient condition of stationarity, i.e. the (\ref{37}) goes to zero when $%
\frac{df(\Psi )}{d\Psi }\neq 0$ consists in the following:

1.\qquad Stream function does not depend explicitly on time.

2.\qquad All singularities do not move.

3.\qquad All dipole moments $D_{i}^{\beta }$ are not function of time.

All terms which contain the velocity $V_{p}$ in equation
(\ref{28}) form obviously the Poisson brackets $\{\Psi ,\Delta
\Psi \}$. That is why the sufficient condition of immobility\ for
all singularities \ and of stationarity for all dipole moments is
the condition when Poisson brackets goes to zero:

\begin{equation}
\{\Psi ,\Delta \Psi \}=0,  \label{38}
\end{equation}

This means that all factors of all independent singularities in Poisson
brackets go to zero (\ref{38}).

\section{Exact stationary solutions with complex singularities}

Now we consider the problem of exact stationary solutions of 2D-Euler
equation, when $\frac{df(\Psi )}{d\Psi }\neq 0.$ Further, it is easy to
consider the Poisson brackets

\begin{equation}
\frac{\partial \Psi }{\partial x}\frac{\partial \Delta \Psi }{\partial y}-%
\frac{\partial \Psi }{\partial y}\frac{\partial \Delta \Psi }{\partial x}=0
\label{39}
\end{equation}

as dimensionless. Let us choose the \textit{anzatz} (\ref{9}) in
the form:

\begin{equation}
\Delta \Psi =\exp \left(-\frac{\Psi }{\Gamma _{0}}\right)+4\pi n\Gamma _{0}\delta (%
\overrightarrow{x}-\overrightarrow{x}_{0})-4\pi \overrightarrow{D}\frac{%
\partial }{\partial \overrightarrow{x}}\delta (\overrightarrow{x}-%
\overrightarrow{x}_{0})  \label{40}
\end{equation}

when function $f(\Psi )$ is chosen in the same way as in the
Stuart work \cite{[4a]} ; we suppose that coefficients $\Gamma $
and $\overrightarrow{D}$ are constant, and the coordinate
$\overrightarrow{x}_{0}$ is not depending on time. (For the sake
of simplicity we can choose $\overrightarrow{x}_{0}=0$). $n-$ is
positive integer number. By means of evident rescaling:

\begin{equation}
\frac{\Psi }{\Gamma _{0}}\rightarrow \Psi ^{^{\prime }},x\rightarrow \Gamma
_{0}^{\frac{1}{2}}x^{^{\prime }},y\rightarrow \Gamma _{0}^{\frac{1}{2}%
}y^{^{\prime }},\frac{\overrightarrow{D}}{\Gamma _{0}^{\frac{3}{2}}}%
\rightarrow \overrightarrow{D}^{^{\prime }}.  \label{41}
\end{equation}

equation (\ref{40}) is reduced to the more simple equation (the
primes were omitted):

\begin{equation}
\Delta \Psi =\exp (-\Psi )+4\pi n\delta (\overrightarrow{x})-4\pi
\overrightarrow{D}\frac{\partial }{\partial \overrightarrow{x}}\delta (%
\overrightarrow{x}).  \label{42}
\end{equation}

(Let us note, that when $\overrightarrow{D}\equiv $0 the equation
(\ref{42}) has the solutions found in \cite{[31]},\cite{[32]}).
First of all, let us find exact solutions for the equation
(\ref{42}), and then we prove, that they are exact stationary
solutions of 2D-Euler equation (\ref{39}). Now we can look for the
solutions of equation (\ref{42}) in the Liouville form:

\begin{equation}
\Psi =-\ln 8\frac{\left| u^{^{\prime }}(z)\right| ^{2}}{(1+\left|
u(z)\right| ^{2})^{2}}.  \label{43}
\end{equation}

when $u^{^{\prime }}(z)=\frac{du(z)}{dz},$ is the unknown for the moment
function of complex variable $z=x+iy$ and $u(z)-$ is primitive function.

Direct calculation of $\Delta \Psi $ (\ref{43}) gives:

\begin{equation}
\Delta \Psi =8\frac{\left| u^{^{\prime }}(z)\right| ^{2}}{(1+\left|
u(z)\right| ^{2})^{2}}-\Delta \ln \left| u^{^{\prime }}(z)\right| ^{2}.
\label{44}
\end{equation}

it is important to note that the formula (\ref{44}) is valid for arbitrary
analytical function $u^{^{\prime }}(z)$ independently of its singularities
structure.

We substitute the formula (\ref{44}) into equation (\ref{42}) and obtain the
equation for the function $\left| u^{^{\prime }}(z)\right| ^{2}:$

\begin{equation}
-\Delta \ln \left| u^{^{\prime }}(z)\right| ^{2}=4\pi n\delta (%
\overrightarrow{x})-4\pi \overrightarrow{D}\frac{\partial }{\partial
\overrightarrow{x}}\delta (\overrightarrow{x}).  \label{45}
\end{equation}

It is easy to see that the equation (\ref{45}) is satisfied, if we choose
the function $\left| u^{^{\prime }}(z)\right| ^{2}$ in the form:

\begin{equation}
\left| u^{^{\prime }}(z)\right| ^{2}=\frac{1}{\left| z\right| ^{2n}}\exp
\overrightarrow{D}\frac{\partial }{\partial \overrightarrow{x}}\ln \left|
z\right| ^{2}.  \label{46}
\end{equation}

Indeed:

\begin{equation}
\ln \left| u^{^{\prime }}(z)\right| ^{2}=-n\ln \left| z\right| ^{2}+%
\overrightarrow{D}\frac{\partial }{\partial \overrightarrow{x}}\ln \left|
z\right| ^{2}.  \label{47}
\end{equation}

The first term in (\ref{47}) gives the Green function of Laplace equation:

\begin{equation}
\Delta \ln \left| z\right| ^{2}=4\pi \delta (\overrightarrow{x})  \label{48}
\end{equation}

and describes the point vortex. The second term in (\ref{47}) is a result of
application of the operator $\overrightarrow{D}\frac{\partial }{\partial
\overrightarrow{x}}$ to the equation (\ref{48}) and describes the point
dipole. Let us introduce the complex dipole moment :

\begin{equation}
D=D_{1}+iD_{2}.  \label{49}
\end{equation}

Then the dipole operator $\overrightarrow{D}\frac{\partial }{\partial
\overrightarrow{x}}$ can be written in the complex form:

\begin{equation}
\overrightarrow{D}\frac{\partial }{\partial \overrightarrow{x}}=D\frac{%
\partial }{\partial z}+\overline{D}\frac{\partial }{\partial \overline{z}}.
\label{50}
\end{equation}

(When $\overline{D}, \overline{z}$ - denote complex conjugated
values).

The function (\ref{46}) can be written down in the form:

\begin{equation}
\left| u^{^{\prime }}(z)\right| ^{2}=\frac{1}{\left| z\right| ^{2n}}\exp \left(D%
\frac{\partial }{\partial z}+\overline{D}\frac{\partial }{\partial \overline{%
z}}\right)(\ln z+\ln \overline{z})=\frac{1}{z^{n}}\exp \left(\frac{D}{z}\right)\frac{1}{%
\overline{z}^{n}}\exp
\left(\frac{\overline{D}}{\overline{z}}\right). \label{51}
\end{equation}

From formula (\ref{51}) follows, that functions $u^{^{\prime
}}(z)$ can be chosen in the form:

\begin{equation}
u^{^{\prime }}(z)=\frac{1}{z^{n}}\exp \left(\frac{D}{z}\right).
\label{52}
\end{equation}

In the point $z=0$, the function $u^{^{\prime }}(z)$ (\ref{52})
has essential singular point, which joins the pole of order $n.$
Now let us find the primitive function $u_{n}(z):$

\begin{equation}
u_{n}(z)=\int \exp \left(\frac{D}{z}\right)\frac{dz}{z^{n}}.
\label{53}
\end{equation}

Using the new variable $w=\frac{D}{z}$, we obtain:

\begin{equation}
u_{n}(w)=-\frac{1}{D^{n-1}}\int W^{(n-2)}\exp WdW.  \label{54}
\end{equation}

From the formula (\ref{54}) we can see that the primitive function
is elementary function only with $n\geq 2.$ This particular case
is examined in this work (others cases will be considered
separately).

Integration by parts in the formula (\ref{54}) with $n\geq 2$,
gives:

\begin{equation}
u_{n}(z)=-\frac{1}{Dz^{n-2}}\exp
\left(\frac{D}{z}\right)P_{n-2}(z); \label{55}
\end{equation}

where the polynomial $P_{n-2}(z)$ has the form:

\begin{eqnarray}
P_{n-2}(z)
&=&1-(n-2)(\frac{z}{D})+(n-2)(n-3)\left(\frac{z}{D}\right)^{2}+\cdots
\label{56} \\
&&+(-1)^{n-3}(n-2)!\left(\frac{z}{D}\right)^{n-3}+(-1)^{n-2}(n-2)!\left(\frac{z}{D}\right)^{n-2}.
\notag
\end{eqnarray}

As a result, $\left| u_{n}(z)\right| ^{2}$ has the form:

\begin{equation}
\left| u_{n}(z)\right| ^{2}=\frac{1}{\left| D\right| ^{2}\left|
z\right| ^{2(n-2)}}\exp
\left(2\frac{D_{1}x+D_{2}y}{x^{2}+y^{2}}\right)\left|
P_{n-2}(z)\right| ^{2}.  \label{57}
\end{equation}

The primitive function $u_{n}(z)$ (\ref{55}), like the function $u^{^{\prime
}}(z)$ (\ref{52}) has in $z=0$ the essential singular point, which joins the
pole of order $(n-2).$ In the real form $\left| u^{^{\prime }}(z)\right|
^{2} $, has obviously the following form:

\begin{equation}
\left| u^{^{\prime }}(z)\right| ^{2}=\frac{1}{(x^{2}+y^{2})^{n}}\exp \left(2\frac{%
D_{1}x+D_{2}y}{x^{2}+y^{2}}\right).  \label{58}
\end{equation}

Consequently, the essential singular point describes in the
complex form the singularities of point dipole kind, while the
pole describes the point vortex, since the expression (\ref{58})
generates following terms in stream function (\ref{43}):

\begin{equation*}
-\ln \left| u^{^{\prime }}(z)\right| ^{2}=n\ln (x^{2}+y^{2})-2\frac{%
D_{1}x+D_{2}y}{x^{2}+y^{2}}.
\end{equation*}

Hence, the exact solution of the equation (\ref{42}) is given by the formula
(\ref{43}), where $u(z)$ is defined by the expression (\ref{55}), while $%
u^{^{\prime }}(z)-$ by the formula (\ref{52}). Now we can prove, that the
obtained solution turns into zero the Poisson brackets (\ref{39}). At first,
let us calculate the velocity field. Derivatives $\frac{\partial \Psi }{%
\partial x}$ and $\frac{\partial \Psi }{\partial y}$ have the form:

\begin{equation*}
\frac{\partial \Psi }{\partial x}=\frac{2}{1+\left| u(z)\right| ^{2}}\frac{%
\partial }{\partial x}u(z)\overline{u(z)},
\end{equation*}

\begin{equation*}
\frac{\partial \Psi }{\partial y}=\frac{2}{1+\left| u(z)\right| ^{2}}\frac{%
\partial }{\partial y}u(z)\overline{u(z)}.
\end{equation*}

Using the formulae:

\begin{equation*}
\frac{\partial }{\partial x}=\frac{\partial }{\partial z}+\frac{\partial }{%
\partial \overline{z}},\ \ \ \ \frac{\partial }{\partial y}=i\left(\frac{\partial }{%
\partial z}-\frac{\partial }{\partial \overline{z}}\right),
\end{equation*}

we obtain a more convenient formula for derivatives:

\begin{equation*}
\frac{\partial \Psi }{\partial x}=\frac{2}{1+\left| u(z)\right| ^{2}}\left(%
\overline{u}\frac{du}{dz}+u\overline{\frac{du}{dz}}\right),
\end{equation*}

\begin{equation*}
\frac{\partial \Psi }{\partial y}=\frac{2i}{1+\left| u(z)\right| ^{2}}\left(%
\overline{u}\frac{du}{dz}-u\overline{\frac{du}{dz}}\right).
\end{equation*}

Taking into account the formula (\ref{52}), after simple algebraic
transformations we obtain the expression for components of
velocity field:

\begin{equation}
\frac{\partial \Psi }{\partial x}=-\frac{2\exp \left(\frac{D\overline{z}+%
\overline{D}z}{\left| z\right| ^{2}}\right)}{(1+\left| u(z)\right|
^{2})\left|
D\right| ^{2}\left| z\right| ^{2n}}(D\overline{z}^{2}\overline{P}_{n-2}+%
\overline{D}z^{2}P_{n-2}),  \label{59}
\end{equation}

\begin{equation}
\frac{\partial \Psi }{\partial y}=-\frac{2i\exp \left(\frac{D\overline{z}+%
\overline{D}z}{\left| z\right| ^{2}}\right)}{(1+\left| u(z)\right|
^{2})\left|
D\right| ^{2}\left| z\right| ^{2n}}(D\overline{z}^{2}\overline{P}_{n-2}-%
\overline{D}z^{2}P_{n-2}).  \label{60}
\end{equation}

(Let us remind, that $n\geq 2$). Now we show that function
(\ref{43}),(\ref {55}) is an exact solution of Poisson brackets
(\ref{39}). For that we
substitute the expression for vorticity (\ref{40}) and derivatives (\ref{59}%
), (\ref{60}) into Poisson brackets (\ref{39}). First of all we
examine the simplest case $n=2.$ In this case the polynomial
$P_{n-2}=1$ and derivatives (\ref{59}), (\ref{60}) take the simple
form:

\begin{equation}
\frac{\partial \Psi }{\partial x}=-\frac{4\exp \left(\frac{D\overline{z}+%
\overline{D}z}{\left| z\right| ^{2}}\right)}{(1+\left| u\right|
^{2})\left| D\right| ^{2}\left| z\right|
^{4}}(D_{1}x^{2}+2D_{2}xy-D_{1}y^{2}), \label{61}
\end{equation}

\begin{equation}
\frac{\partial \Psi }{\partial y}=\frac{4\exp \left(\frac{D\overline{z}+\overline{%
D}z}{\left| z\right| ^{2}}\right)}{(1+\left| u\right| ^{2})\left|
D\right| ^{2}\left| z\right|
^{4}}(D_{2}x^{2}-2D_{1}xy-D_{2}y^{2}),  \label{62}
\end{equation}

\begin{equation}
\left| u(z)\right| ^{2}=\frac{1}{\left| D\right| ^{2}}\exp \left(\frac{D\overline{%
z}+\overline{D}z}{\left| z\right| ^{2}}\right).  \label{63}
\end{equation}

We write the Poisson brackets (\ref{39}) in the explicit form:

\begin{equation}
\{\Psi ,\Delta \Psi \}=4\pi n\left[\frac{\partial \Psi }{\partial
x}\delta (x)\delta ^{^{\prime }}(y)-\frac{\partial \Psi }{\partial
y}\delta (y)\delta ^{^{\prime }}(x)\right]-  \label{64}
\end{equation}

\begin{equation*}
-4\pi \left [ D_{1}\frac{\partial \Psi }{\partial
x}-D_{2}\frac{\partial \Psi }{\partial y}\right]\delta ^{^{\prime
}}(x)\delta ^{^{\prime }}(y)-
\end{equation*}

\begin{equation*}
-4\pi \left [ D_{2}\frac{\partial \Psi }{\partial x}\delta
(x)\delta ^{^{\prime \prime }}(y)-D_{1}\frac{\partial \Psi
}{\partial y}\delta (y)\delta ^{^{\prime \prime }}(x)\right].
\end{equation*}

It is obvious, that all the terms in the first brackets (\ref{64}) are equal
to zero because they contain this kind of zeros:

\begin{equation*}
x^{2}\delta (x),x\delta (x),x^{2}\delta ^{^{\prime }}(x); \ \ \ \
\ y^{2}\delta (y),y\delta (y),y^{2}\delta ^{^{\prime }}(y).
\end{equation*}

In the second brackets one part of terms is also equal to zero, but there
are dangerous terms of this type: $xy\delta ^{^{\prime }}(x)\delta
^{^{\prime }}(y).$ However, these terms are part of second brackets in the
following combination:

\begin{equation*}
\left(D_{1}\frac{\partial \Psi }{\partial x}-D_{2}\frac{\partial \Psi }{\partial y%
}\right)\delta ^{^{\prime }}(x)\delta ^{^{\prime }}(y)=[\cdots
](-D_{1}D_{2}xy+D_{2}D_{1}xy)\delta ^{^{\prime }}(x)\delta
^{^{\prime }}(y)=0,
\end{equation*}

i.e. are reciprocally cancelled. (Here brackets $[\cdots ]$ denote
the common factor). Now we consider the last brackets in
(\ref{64}). In this brackets also, one part of terms turns into
zero at once, but there are dangerous
terms of this kind: $y^{2}\delta (x)\delta ^{^{\prime \prime }}(y)$ and $%
x^{2}\delta (y)\delta ^{^{\prime \prime }}(x)$. These dangerous terms are
part of the brackets (\ref{64}) in the following combination:

\begin{equation}
\lbrack \cdots ][D_{2}D_{1}y^{2}\delta (x)\delta ^{^{\prime \prime
}}(y)-D_{1}D_{2}x^{2}\delta (y)\delta ^{^{\prime \prime }}(x)]  \label{65}
\end{equation}

(Here brackets $[\cdots ]$ denote the common factor). Now we use
the formula (\ref{A.4}). From this formula one can see, that
dangerous terms have the form:

\begin{equation}
D_{1}D_{2}\delta (x)\delta (y)\frac{d^{2}}{dy^{2}}y^{2}-D_{1}D_{2}\delta
(x)\delta (y)\frac{d^{2}}{dx^{2}}x^{2}=0  \label{66}
\end{equation}

and are cancelled in commutator (\ref{64}). Others terms are obviously zero.
Consequently, the Poisson brackets turn into zero for all singularities. \
According to the results of chapter 3, this guarantees that singularities do
not move and that the dipole moment $\overrightarrow{D}$ is conserved. It is
proved that the obtained solution of equation (\ref{42}), is exact,
stationary solution of 2D-Euler equation (\ref{39}) with $n=2.$ let us
consider now the general case $n\succ 2.$ In this case velocities (\ref{59}%
), (\ref{60}) contain the polynomials $P_{n-2}(z)$ i $\overline{P_{n-2}}(z)$
(\ref{56}).

It is \ clear now that the additional powers $z$ or $\overline{z}$ in these
polynomials generate in Poisson brackets zero terms only. The dangerous
terms coincide only with the first term in the polynomial $P_{n-2}(z)$, i.e.
unit. But these \ terms correspond to the case $n=2$ , and are already been
considered. Hence, we prove that formulae in (\ref{43}), together with the
function $u_{n}(z)$ (\ref{55}), give exact stationary solution of 2D-Euler
equation with $n\geq 2.$ In explicit form this solution has the form :

\begin{equation}
\Psi =-\ln 8-\ln \left| u^{^{\prime }}(z)\right| ^{2}+2\ln (1+\left|
u(z)\right| ^{2})=  \label{67}
\end{equation}

\begin{equation*}
=-\ln 8+n\ln (x^{2}+y^{2})-2\frac{D_{1}x+D_{2}y}{x^{2}+y^{2}}+
\end{equation*}

\begin{equation*}
+2\ln \left[1+\frac{\left| P_{n-2}(z)\right| ^{2}}{\left| D\right|
^{2}(x^{2}+y^{2})^{(n-2)}}\exp
\left(2\frac{D_{1}x+D_{2}y}{x^{2}+y^{2}}\right)\right].
\end{equation*}

From equation (\ref{67}) it follows that the essential singularity
splits up into singularities of point vortex and point dipoles
types. However, as we will see later, the fusion of these
singularities leads to a singular point with more complex geometry
with a vector field index equal to three. This can be interpreted
as the sum of the indexes of the point vortex and the point
dipole.

\section{Examples of vortex structures with complex singularities}

\bigskip 1) First of all let us examine the simplest case, when $n=2.$ In
this case the function $u_{2}(z)$ (\ref{55}) has the form:

\begin{equation}
u_{2}(z)=-\frac{1}{D}\exp \left(\frac{D}{z}\right),  \label{68}
\end{equation}

The function $\left| u_{2}(z)\right| ^{2}$ is given by (\ref{63}), $%
u_{2}^{^{\prime }}(z)$ has the form:

\begin{equation}
\left| u_{2}^{^{\prime }}(z)\right| ^{2}=(x^{2}+y^{2})^{-2}\exp \left(2\frac{%
D_{1}x+D_{2}y}{x^{2}+y^{2}}\right).  \label{69}
\end{equation}

As a result we find the stream function (\ref{43}):

\begin{equation}
\Psi =-\ln 8+2\ln (x^{2}+y^{2})-2\frac{D_{1}x+D_{2}y}{x^{2}+y^{2}}+
\label{70}
\end{equation}

\begin{equation*}
+2\ln \left[1+\frac{1}{D_{1}^{2}+D_{2}^{2}}\exp \left(2\frac{D_{1}x+D_{2}y}{x^{2}+y^{2}%
}\right)\right]
\end{equation*}

\begin{figure}[h]
  \centering
  \includegraphics[width=6 cm]{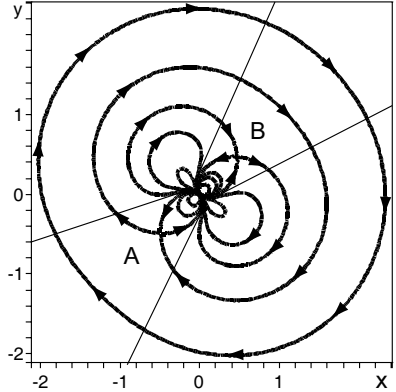}\\
  \caption{The simplest vortex structure with the index
  of vector field equal to 3 (n=2). Complex singularity is a result
   of  the fusion of essential singular point (dipole singularity)
    with the pole  (point vortex). One can see the presence of two
    hyperbolic points A and B, external and internal correspondingly.
    Separatrix of these points connect these hyperbolic points  with
    the central complex singular point. Straight lines indicate exceptional
     directions of central singular point. }\label{Fig3}
\end{figure}

Farther, we can consider that $D_{1}=D_{2}.$ The vortex structure which is
described by the stream function (\ref{70}), is presented on (Fig.\ref{Fig3}), with $%
D_{1}=D_{2}=1.$ It is clear, that with great values $\left| \overrightarrow{x%
}\right| ^{2}\rightarrow \infty ,$ the stream function (\ref{70})
coincides asymptotically with the stream function of point vortex
with negative vorticity. From (Fig.\ref{Fig3}) one can see that
inside of external closed stream line there is a vortex structure
with the a non trivial topology of stream line. In the center
there is a complex singular point, which results from the
coincidence of singularities of point vortex and point dipole
types.
Furthermore, one can see the presence of two hyperbolic points, one outside $%
A$ and another inside $B$ . Separatrix of these points link the
hyperbolic points with the central complex singular point. We
examine more in details the structure of complex singular point.
The general theory of such kind of singularities is stated in
qualitative theory of ordinary differential equations (see, for
example, \cite{[49]}, \cite{[50]} ). According to this theory,
first of all, we need to choose exceptional directions of the
singular point. This are directions of tangents, following which
the infinite number of integral curves go inside of the singular
point and outside of it. One can see from (Fig.\ref{Fig3}), that
there are four such exceptional directions which are designated on
(Fig.\ref{Fig3}) by direct lines. Integral curves which are inside
of exceptional lines form sectors. In our case, between the lines
there are only four elliptical sectors (see, for example,
\cite{[50]} ). According to the general theory, the index $J(0)$
of complex singular point is given by Bendixson's formula (see,
for example, \cite{[49]}):

\begin{equation}
J(0)=\frac{1}{2}(2+n_{e}-n_{h}),  \label{71}
\end{equation}

where $n_{e}-$ is the number of elliptic sectors and $n_{h}-$ is the number
of hyperbolic sectors. In our case $n_{h}=0,n_{e}=4.$ That is why

\begin{equation}
J(0)=3.  \label{72}
\end{equation}

The condition (\ref{72}) means, that the complex singular point is
structurally stable, because the necessary and sufficient condition of the
structural stability for complex singular point on plane is, that its index $%
J(0)\neq 0.$ (see, for example, \cite{[50]}). Note, that in case
of dipole singularity exceptional directions coincide with vector
direction of dipole moment $\overrightarrow{D},$ while dipole
index $J_{D}=2,$ i.e. point dipole is the structurally stable
singularity. Index of complex singular point can be found
otherwise without using of the general theory. For this, let us
cut out by circles singular points as it is shown on
(Fig.\ref{Fig4}). Then, in the obtained multi connected domain the
index of vector field is equal to zero. I.e.:

\begin{figure}
  \centering
  \includegraphics[width=6 cm]{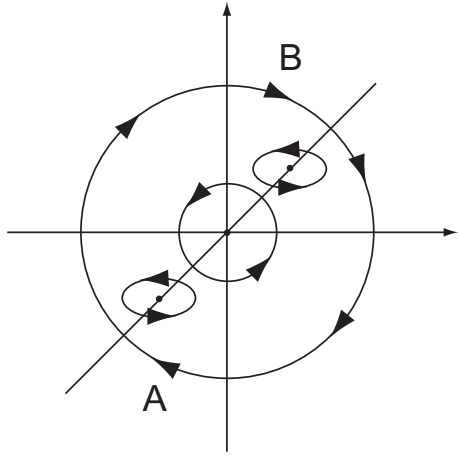}\\
  \caption{Contour, which is used to find the index of complex singular point.}\label{Fig4}
\end{figure}

\begin{equation}
\sum_{i}J_{i}+J(0)+J_{s}=0,  \label{73}
\end{equation}

where $J_{s}-$ is the index of outside circle $S$,
$\sum_{i}J_{i}-$ is the sum of indices of all internal simple
singular points. This means, that the index of vortex structure,
which is surrounded by contour $S$, is equal to:

\begin{equation}
J_{A}+J_{B}+J(0)=1.  \label{74}
\end{equation}

Since index of hyperbolic points $A$ and $B$ is equal to $(-1)$,
then the equation (\ref{74}), gives index of complex singular
point $J(0)=+3$.

2) Now, let us examine the case $n=3$ (dipole plus pole of order
$n=3$). In this case the polynomial $P_{n-2}(z)$ is not trivial:
$P_{1}(z)=1-\frac{z}{D}$.

Function $\left| u_{3}(z)\right| ^{2}$ has the form :

\begin{equation}
\left| u_{3}(z)\right| ^{2}=\frac{1}{\left| D\right| ^{4}}\left| \frac{D}{z}%
\right| ^{2}\exp \left(\frac{D\overline{z}+\overline{D}z}{\left| z\right| ^{2}}%
\right)\left| 1-\frac{z}{D}\right| ^{2},  \label{75}
\end{equation}

or in the real form:

\begin{equation}
\left| u_{3}(z)\right|
^{2}=\frac{1}{(x^{2}+y^{2})(D_{1}^{2}+D_{2}^{2})}\exp
\left(2\frac{D_{1}x+D_{2}y}{x^{2}+y^{2}}\right)\times  \label{76}
\end{equation}

\begin{equation*}
\times \left [ 1+\frac{(x^{2}+y^{2})}{(D_{1}^{2}+D_{2}^{2})}-2\frac{%
(D_{1}x+D_{2}y)}{(D_{1}^{2}+D_{2}^{2})}\right].
\end{equation*}

\begin{figure}
  \centering
  \includegraphics[width=6 cm]{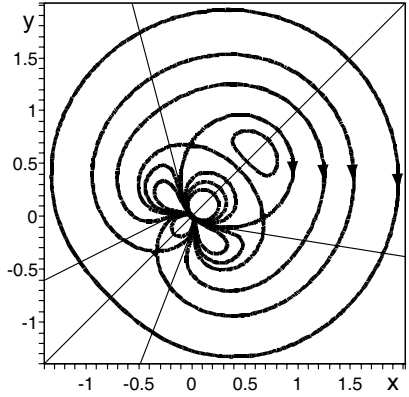}\\
  \caption{Vortex structure with n=3 (dipole and pole of order n=3).
   One can see, that the elliptic point appears. Internal singular
   hyperbolic point splits into two. Structure of central singular
   point does not change. Straight lines indicate, as earlier, exceptional
    directions of central singular point. Structure of the external
    separatrix does not change. Structure of internal separatrix which are
     now surrounding new elliptic points becomes more complex. }\label{Fig5}
\end{figure}

Correspondingly the \ stream function $\Psi $ has the form:

\begin{equation}
\Psi =-\ln 8+3\ln (x^{2}+y^{2})-2\frac{D_{1}x+D_{2}y}{x^{2}+y^{2}}+
\label{77}
\end{equation}

\begin{equation*}
+2\ln [1+\left| u_{3}(z)\right| ^{2}],
\end{equation*}

where $\left| u_{3}(z)\right| ^{2}$ is given by the formula
(\ref{76}). Stream lines picture is presented on (Fig.\ref{Fig5})
with $D_{1}=D_{2}=1.$ One can see that, unlike previous case, the
elliptical point appears in the solution. Previous internal
hyperbolic point splits into two. Structure of central singular
point does not change, its index is still equal to 3.

Note, that outside separatrix remains the same. It surrounds all \ internal
vortex structure, including new elliptical point around which appears the
vortex with negative vorticity. The separatrix of previous hyperbolic
internal point gets more complex because now it surrounds another additional
elliptical point. All separatrix link hyperbolic points either with each
other or with central singularity.

3) Now we consider the case $n=4$ (dipole plus pole of the order
$n=4$). In this case the polynomial $P_{n-2}(z)$ has the form:

\begin{equation}
P_{2}(z)=1-2\left(\frac{z}{D}\right)+2\left(\frac{z}{D}\right)^{2}.
\label{78}
\end{equation}

In the real form:

\begin{equation}
\left| P_{2}(z)\right| ^{2}=1-4\frac{D_{1}x+D_{2}y}{(D_{1}^{2}+D_{2}^{2})}%
\left(1+2\frac{(x^{2}+y^{2})}{(D_{1}^{2}+D_{2}^{2})}\right)+
\label{79}
\end{equation}

\begin{equation*}
+4\frac{[(D_{1}^{2}-D_{2}^{2})(x^{2}-y^{2})+4D_{1}D_{2}xy]}{%
(D_{1}^{2}+D_{2}^{2})^{2}}+4\frac{(x^{2}+y^{2})}{(D_{1}^{2}+D_{2}^{2})}\left(1+%
\frac{(x^{2}+y^{2})}{(D_{1}^{2}+D_{2}^{2})}\right).
\end{equation*}

Function $\left| u_{4}(z)\right| ^{2}$ takes the form:

\begin{equation}
\left| u_{4}(z)\right| ^{2}=\frac{\left| P_{2}(z)\right| ^{2}}{\left|
D\right| ^{2}\left| z\right| ^{4}}\exp \left(\frac{D\overline{z}+\overline{D}z}{%
\left| z\right| ^{2}}\right).  \label{80}
\end{equation}

As a result for stream function we obtain the expression:

\begin{equation}
\Psi =-\ln 8+4\ln (x^{2}+y^{2})-2\frac{D_{1}x+D_{2}y}{x^{2}+y^{2}}+
\label{81}
\end{equation}

\begin{equation*}
+2\ln [1+\left| u_{4}(z)\right| ^{2}].
\end{equation*}

\begin{figure}
  \centering
  \includegraphics[width=6 cm]{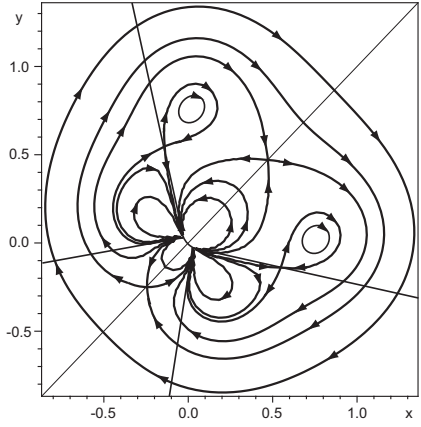}\\
  \caption{Vortex structure with n=4, "butterfly"(dipole and pole of order n=4).
   One can see, that two elliptic singular points appear with two vortices
    satellites. Besides, one new group of separatrix appears, whiich connects
    vortices satellites with the central singular point.Topological  structure
     of separatrix  of external and internal hyperbolic points does not change.}\label{Fig6}
\end{figure}

The image of stream lines is presented on (Fig.\ref{Fig6}), with
$D_{1}=D_{2}=1.$ (Butterfly). First of all one can see that two
singular elliptical points appear and around them two satellites
vortices. We can see also that there are four hyperbolic points.
Separatrix structure becomes more complex. There are three groups
of separatrix. First separatrix is of the same kind that outside
separatrix in all previous cases. The second one is the same as
the separatrix of internal hyperbolic point in first case. So a
new group of separatrix appears which links vortices satellites
(elliptical singular points) with the central singular point; the
structure of the last one does not change from tropological point
of view.

4). Let us examine poles of higher order :$n=5$, i $n=6.$
Correspondingly polynomials $P_{n-2}(z)$ have the form:

\begin{equation}
P_{3}(z)=1-3\left(\frac{z}{D}\right)+6\left(\frac{z}{D}\right)^{2}-6\left(\frac{z}{D}\right)^{3},
\label{82}
\end{equation}

\begin{equation}
P_{4}(z)=1-4\left(\frac{z}{D}\right)+12\left(\frac{z}{D}\right)^{2}-24\left(\frac{z}{D}\right)^{3}
+24\left(\frac{z}{D}\right)^{4}.  \label{83}
\end{equation}

\begin{figure}
  \centering
  \includegraphics[width=6 cm]{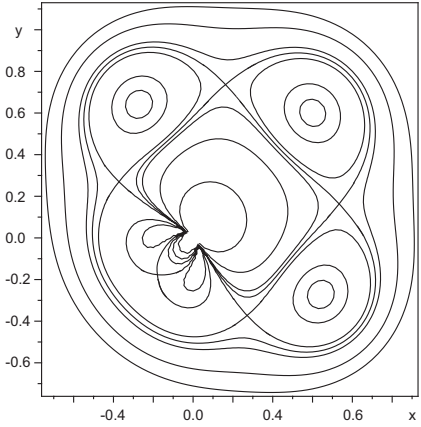}\\
  \caption{Vortex structure with n=5. In this case three elliptic singular
  points  appear (three vortices satellites) and 5 hyperbolic points.}\label{Fig7}
\end{figure}

Stream lines with $n=5$ are presented on (Fig.\ref{Fig7}), with
$D_{1}=D_{2}=1.$ In this case three elliptical singular points
appear (three vortices satellites) and five hyperbolic singular
points. In case of $n=6$, four elliptical singular points appear
and six hyperbolic singular points. With the increasing of number
$n$ there are always $n$- hyperbolic singular points and $n-2$
elliptical points. Central singular point conserves four
exceptional directions, i.e. its index remains equal to 3.
Appeared vortex structure has symmetry relative to square
diagonal. Diagonal pass always by central singular points and by
opposite singular point, which is hyperbolic with $n=2k,$ and
elliptical with $n=2k+1.$ The obtained vortex structure has the
form of necklace composed of vortices satellites except the low
sector, which always has hyperbolic singular point, linked by
separatrix with central singularity.

\section{Discussions and conclusions}

In this work we want to call attention to the fact that there are exact
solutions of 2D-Euler equation which contain point singularities more
complex, than singularities which are usually considered as typical ones for
2D-Euler equation. Such complex singularities can be non stationary \cite
{[47]}, or stationary as well. Let us remind, that complex singularities are
defined as singular points of vector field, which have the index $\left|
J\right| \geq 2$ . The simplest singularity of this type is the dipole one
with the index $J=2$. Point vortices and dipole singularities form a set of
moving singularities in 2D-Euler equation, which dynamics is Hamiltonian
\cite{[47]}. With $f(\Psi )=0$, for general case, moving singularities can
be only point vortices and \ point dipoles. It means, that its index can not
exceed two. \ The reason for this, as it was shown earlier, that there are
self-interaction of multipoles and overdetermination of their motion
equations.

Now let us examine the case when in expansion (\ref{9}) the function $f(\Psi
)\neq 0.$ Consider more specifically the anzatz (\ref{9}) in the form:

\begin{equation}
\Delta \Psi =f(\Psi )-\Gamma _{0}\delta (\overrightarrow{x}-\overrightarrow{x%
}_{0})-\overrightarrow{D}\frac{\partial }{\partial \overrightarrow{x}}\delta
(\overrightarrow{x}-\overrightarrow{x}_{0}).  \label{5.3}
\end{equation}

It is not difficult to see, that with $f(\Psi )=0$, there is no
stationary solutions for 2D-Euler equation because $\left\{ \Delta
\Psi ,\Psi \right\} \neq 0$. If the function $f(\Psi )\neq 0$\ and
is chosen in the Stuart's form:

\begin{equation}
f(\Psi )=\exp (-\Psi ),  \label{5.4}
\end{equation}

the situation change substantially, what can be seen from results of this
work. Presence of smooth part of vorticity field in the equation (\ref{5.3})
gives exact solutions with the more complex singularity of index 3. As it is
shown in this work, the singularity of vector field of index 3 can be
interpreted in complex form as a fusion in function $u^{^{\prime }}(z)$ (\ref
{52}) of the pole, which corresponds to point vortex, with essential
singular point which corresponds to point dipole. In real form, as it can be
seen from the equation (\ref{67}), for the final stream function the
essential singularity splits up into singularities of point vortex and point
dipoles types. However, the fusion of these singularities leads to a
singular point with more complex geometry with a vector field index equal to
three. This can be interpreted as the sum of the indexes of the point vortex
and the point dipole.Exact localized solutions obtained in this work
describe vortex structure of the complex form, where the singular point is
surrounded be vortices satellites. With increasing of the number $n$ \ the
vortices satellites have tendency to form symmetrical necklaces. The
existence of \ exact solutions with complex singularities itself is an
important fact that is why in this work we contented ourself with
consideration of simplest class of exact solutions expressed in elementary
functions. We did not deal with questions of the construction of more
complex solutions expressed by special functions and with important
questions of stability of vortex configurations with complex singular
points. All these questions must be examined separately and some of them
will be studied in next works.

\section{Appendix. Action of derivatives of $\protect\delta -$function on
velocity field}

For convenience we present in this Appendix some formulae used in this work.
It is well known (see, for example, \cite{[52a]} ), that generalized
functions are linear functionals, which are acting in the space of basic
functions $\{\varphi (x)\}.$ By definition, the \ derivative $\delta
^{^{\prime }}(x)$ acts on differentiable function $a(x)$ according to
formula:

\begin{equation}
\int \delta ^{^{\prime }}(x)a(x)\varphi (x)dx=-\int \delta (x)a^{^{\prime
}}(0)\varphi (x)dx-\int \delta (x)a(0)\varphi ^{^{\prime }}(x)dx=
\label{A.1}
\end{equation}

\begin{equation*}
=-\int \delta (x)a^{^{\prime }}(0)\varphi (x)dx+\int \delta ^{^{\prime
}}(x)a(0)\varphi (x)dx;\forall \varphi (x).
\end{equation*}

This gives a well known formula:

\begin{equation}
a(x)\delta ^{^{\prime }}(x)=-a^{^{\prime }}(x)\delta (x)+a(0)\delta
^{^{\prime }}(x)  \label{A.2}
\end{equation}

Derivative of order $(k)$ acts on infinitely differentiable function $a(x)$,
according to formula (\ref{A.3}), which is obtained in a similar manner (\ref
{A.1}), as a result of integration by parts:

\begin{equation}
a(x)\delta ^{(k)}(x)=\sum_{j=0}^{k}(-1)^{j+k}C_{k}^{j}a^{(k-j)}(0)\delta
^{(j)}(x).  \label{A.3}
\end{equation}

We need to use now particular cases: $k=2$:

\begin{equation}
a(x)\delta ^{^{\prime \prime }}(x)=a^{^{\prime \prime }}(0)\delta
(x)-2a^{^{\prime }}(0)\delta ^{^{\prime }}(x)+a(0)\delta ^{^{^{\prime \prime
}}}(x),  \label{A.4}
\end{equation}

and $k=3:$

\begin{equation}
a(x)\delta ^{^{^{\prime \prime \prime }}}(x)=-a^{^{\prime \prime \prime
}}(0)\delta (x)+3a^{^{\prime \prime }}(0)\delta ^{^{\prime
}}(x)-3a^{^{\prime }}(0)\delta ^{^{\prime \prime }}(x)+a(0)\delta
^{^{^{\prime \prime \prime }}}(x).  \label{A.4.1}
\end{equation}

From formula (\ref{A.3}) one can see that the derivative of order
$(k)$ from $\delta -$ function engenders also all derivatives of
lower orders including terms without derivatives, i.e.simply
$\delta -$ functions. Similar formulae are obtained also for the
case of multiple variables. In particular, for two variables
$x_{i_{1}}, x_{i_{2}}$ the second derivative acts according the
formula:

\begin{equation}
a(\overrightarrow{x})\frac{\partial ^{2}}{\partial x_{i_{1}}\partial
x_{i_{2}}}\delta (\overrightarrow{x})=\left[\frac{\partial ^{2}a(\overrightarrow{x%
})}{\partial x_{i_{1}}\partial x_{i_{2}}}\mid
_{\overrightarrow{x}=0}\right]\delta
(\overrightarrow{x})-\left[\frac{\partial
a(\overrightarrow{x})}{\partial
x_{i_{2}}}\mid _{\overrightarrow{x}=0}\right]\frac{\partial \delta (%
\overrightarrow{x})}{\partial x_{i_{1}}}-  \label{A.5}
\end{equation}

\begin{equation*}
-\left[\frac{\partial a(\overrightarrow{x})}{\partial x_{i_{1}}}\mid _{%
\overrightarrow{x}=0}\right]\frac{\partial \delta
(\overrightarrow{x})}{\partial x_{i_{2}}}+a(0)\frac{\partial
^{2}\delta (\overrightarrow{x})}{\partial x_{i_{1}}\partial
x_{i_{2}}}
\end{equation*}


\bigskip

\begin{thebibliography}{99}
\bibitem{[1]}  H.Lamb, \textit{Hydrodynamics}, 6th ed.Cambridge University
Press, Cambridg,1932.

\bibitem{[2]}  P.G.Saffman, \textit{Vortex Dynamics, }Cambridge University
Press, Cambridge,1992.

\bibitem{[3]}  P.K.Newton, \textit{The N-vortex problem: analytical
techniques, }Springer Verlag, N.Y., 2001.

\bibitem{[4a]}  J.T.Stuart,J.Fluid Mech.29,417,(1967).

\bibitem{[5]}  T.B.Mitchell and L.F.Rossi,Phys. Fluids,20,054103 (2008).

\bibitem{[6]}  W.Kramer,H.J.H.Clercx, and G.J.F.van Heijst,Phys.
Fluids,19,126603 (2007).

\bibitem{[7]}  B.Legras and D.G.Dritschel,,Phys.Fluids.A3,845,(1991).

\bibitem{[8]}  M.V.Melander,A.S.Styczek,N.J.Zabusky,Phys.Rev.Lett.,
53,p.1222,(1984).

\bibitem{[9]}  N.J.Zabusky,Physica D18,N1/3,p.15,(1986).

\bibitem{[10]}  G.Dritschel and B.Legras, Phys.Today,46(3),44 (1993)

\bibitem{[11]}  D.A.Schecter,K.S.Fine,D.H.E.Dubin, and
C.F.Driscoll,Phys.Fluids.11,905,(1999).

\bibitem{[12]}  H.Aref,P.K.Newton,M.A.Stremler,T.Tokieda,and
D.I..Vainchtein,Adv.Appl.Mech.39,1,(2003).

\bibitem{[13]}  Z.J. Dezhe and D.H.E.Dubin, Phys.Fluids,v.13,N3,
p.677,(2001).

\bibitem{[14]}  D.G.Dritschel,J.FluidMech.172,p.157,(1986).

\bibitem{[15]}  J.C.Mc Williams,J.Fluid.Mech.,146,p.21,(1984).

\bibitem{[16]}  I.Couder,C.Basdevadt,J.Fluid Mech.,173,p.225,(1986).

\bibitem{[17]}  J.Sommeria,S.P.Meyers, and H.L.Swinney ,
Nature,London,331,689,(1988).

\bibitem{[18]}  Y.G.Morel and X.J.Carton,J.Fluid Mech.267,23,(1994)

\bibitem{[19]}  G.I.F. Van Heijst and J.B.Flor,Nature,London,340,212,(1989).

\bibitem{[20]}  X.J.Carton and B.Legras,J.Fluid Mech.,267,53,(1994).

\bibitem{[21]}  L.A.Barba and A.Leonard,Phys.Fluids,19,017101,(2007).

\bibitem{[22]}  R.Mallier and S.A.Maslowe,Phys.Fluids,5,1074,(1994).

\bibitem{[23]}  K.W.Chow,N.W.M.Ko,R.C.K.Leung and S.K.Tang, Phys.Fluids
v.10,n5,p.1111 (1998).

\bibitem{[24]}  D.Gurarie,K.W.Chow, Phys.Fluids v.16,n9,p.3296, (2004).

\bibitem{[25]}  A.A.Abrashin and E.I.Yakubovich,Sov.Phys.Dokl.29,370,(1984).

\bibitem{[26]}  S.KIda, J.Phys.Soc.Japan 50,3517,(1981)

\bibitem{[27]}  J.Neu,Phys.Fluids 27,(10),2397,(1984).

\bibitem{[28]}  D.Crowdy,Phys.Fluids,v.11,n9,p.2556,(1999).

\bibitem{[29]}  D.Crowdy,Phys.Fluids,14,(1),p.257,(2002).

\bibitem{[30]}  D.Crowdy,J.Fluid Mech.,v.469,209,(2002).

\bibitem{[31]}  D.Crowdy,Phys.Fluids,v.15,n.12,p.3710,(2003).

\bibitem{[32]}  A.Tur, and V.Yanovsky,Phys.Fluids,v.16,n.8,p.2877,(2004).

\bibitem{[33]}  V.V.Meleshko and M.Y.Konstantinov, \textit{Vortex dynamics
and chaotic phenomena,}World Scientific, Singapor,1999.

\bibitem{[34]}  H.Aref, Annu.Rev.Fluid Mech.,1986,173.

\bibitem{[35]}  P.Boyland,M.Stremler, and H.Aref,Physica D 175,69,(2003).

\bibitem{[36]}  B.Eckhart and H.Aref,Philos.Trans.R.Soc.London
Ser.A326,655,(1988).

\bibitem{[37]}  H.Aref,S.W.Jones,S.Mofina,I.Zavadsky,Physica D 37,423,(1989).

\bibitem{[38]}  H.Aref,N.Rott,H.Thomann,Ann.Rev.Fluid Mech. 24,1,(1992).

\bibitem{[39]}  P.K.Newton,and H.Shokraneh,Proc.R.Soc.A,462,149 (2006).

\bibitem{[40]}  H.Marmanis,Proc.R.Soc.Lond.A,464,586 (1998).

\bibitem{[41]}  P.K.Newton,and H.Shokraneh,Proc.R.Soc.A,464,1525, (2008).

\bibitem{[42]}  G.R.Flierl,V.D.Larichev,J.C.McWilliams,and
G.M.Reznik,Dyn.Atmos.Oseans 5,1,(1980)

\bibitem{[43]}  D.D.Hobson,Phys.Fluids,A3 ,(12),p.3027,(1991).

\bibitem{[44]}  X.N.Su,W.Horton,and P.J.Morrison,Phys.Fluids,B3
,(4),p.921,(1991).

\bibitem{[45]}  C.Matsuoka, and K.Nosaki,Phys.Fluids,B4 ,(3),p.551,(1992).

\bibitem{[46]}  M.Kono,and W.Horton,Phys.Fluids,B3 ,(12),p.3255,(1991).

\bibitem{[46a]}  %
O.G.V.Fuentes,andG.J.F.vanHeijst,Phys.Fluids,7(11),p.2735,(1995).

\bibitem{[47]}  V.Yanovsky,A.Tur,and K.Kulik,Phys.Lett.A 373,2484, (2009).

\bibitem{[48]}  K.Kulik,A.Tur,V.Yanovsky,Theor.Mat.Phys.162,383,(2010).

\bibitem{[49]}  P.Hartman, \textit{Ordinary differetial equations},
J.Wiley\&Sons,N.Y.1964.

\bibitem{[50]}  V.V.Nemitsky, and V.V.Stepanov,\textit{Qualitative theory of
differential equations, }Princeton,N.J. 1960.

\bibitem{[52a]}  I.M.Gelfand, and G.E.Shilov, \textit{Generalized functions}, Academic
Press,N.Y.London,1967.
\end{thebibliography}
\end{document}